\documentstyle[12pt]{article}
\catcode `@=11
\@addtoreset{equation}{section}

\newdimen\z@ \z@=0pt
\def\m@th{\mathsurround=\z@}
\def\ialign{\everycr{}\tabskip\z@skip\halign} 
\def\eqalign#1{\null\,\vcenter{\openup\jot\m@th
  \ialign{\strut\hfil$\displaystyle{##}$&$\displaystyle{{}##}$\hfil
      \crcr#1\crcr}}\,}
\def\matrix#1{\null\,\vcenter{\normalbaselines\m@th
    \ialign{\hfil$##$\hfil&&\quad\hfil$##$\hfil\crcr
      \mathstrut\crcr\noalign{\kern-\baselineskip}
      #1\crcr\mathstrut\crcr\noalign{\kern-\baselineskip}}}\,}
\catcode `@=12
\def\frac#1#2{{#1\over#2}}

\def\be{\begin{equation}}
\def\ee{\end{equation}}

\begin{document}

\begin{center}
\begin{Large}
\bf{Free differential algebras and generic 2D dilatonic (super)gravities}
\footnote{Work partially supported by the 
  {\it DGICYT} (Spain) under contract  PB96-0756.}
\\[0.5cm]
\end{Large}
\begin{large}
J.~M.~Izquierdo
\end{large}
\\[0.3cm]
\begin{it}
 Departamento de F\'{\i}sica Te\'orica, Universidad de Valladolid\\
E--47011, Valladolid, Spain
\end{it}
\end{center}

\begin{abstract}
The field equations for both generic bosonic and generic locally 
supersymmetric 2D dilatonic gravity theories in the absence of matter
are written as free differential algebras. This constitutes a 
generalization of the gauge theoretic formulation. Moreover, it is shown 
that the condition of free differential algebra can be
used to obtain the equations in the locally supersymmetric case. 
Using this formulation, the general solution of the
field equations is found in the language of differential forms. 
The relation with the ordinary formulation and the coupling to 
supersymmetric conformal matter are also studied.
\end{abstract}

\section{Introduction}
\label{sec2}

 Low-dimensional models of gravity  play an important
role in present-day theoretical physics due to the complexity of 
four-dimensional gravity. They are used to investigate the consequences
of general covariance or local supersymmetry in a simpler setting.
In two dimensions, although pure gravity is trivial, there are some models
that include an additional field which have some properties analogous
to the ones  arising in four-dimensional general relativity,
a fact that may be used to try to answer some fundamental problems
also present in the four-dimensional case. For
instance, black hole formation and evaporation can be studied using
dilatonic type models of two-dimensional gravity such as the 
string-inspired dilatonic gravity (also known as 
CGHS model \cite{Cal:Gid:Har:Str:92}).

The CGHS model is not the only one that has been studied. Other interesting 
models are the Jackiw-Teitelboim model \cite{Tei:85} and the more
realistic, although less simple, spherically symmetric 
 reduction of four-dimensional gravity. In fact, 
by using appropriate conformal redefinitions of the metric, it is
always possible to write the action of a generic first-order model  of 2D 
dilaton gravity conformally coupled to matter in the form \cite{Ban:OLa:91} 
\be
      S=\int d^2x\sqrt{-g}\left[ R\eta+V(\eta)-\frac{1}{2}\partial^\mu f
  \partial_\mu f\right]\quad ,                           \label{generic}
\ee
where $f$ is a matter field (only one is considered for simplicity), and 
$V(\eta)$ is an arbitrary potential term.
The  CGHS model is obtained when $V=4\lambda^2$ (for constant 
$\lambda$), the Jackiw-Teitelboim model corresponds to 
$V=4\lambda^2\eta$ and for the spherically symmetric
 reduction of four dimensional gravity   $V\propto
\frac{1}{\sqrt{\eta}}$. The case of 
the model with an exponential potential ($V=4\lambda^2
e^{\beta\eta}$, for constant $\beta$) has also been considered \cite{Mann:94}.

 The study of
the consequences of general covariance in two-dimensional theories has
been extended to the locally supersymmetric case. This gives 
information not only about supergravity  itself, but also about the
consequences its presence may have in the bosonic part of the 
theory, such as positivity of the energy 
\cite{Par:Str:93,Bil:93}. The generalization
of (\ref{generic}) to the locally supersymmetric case was given in
\cite{Par:Str:93} using the superfield formulation of 
\cite{How:79}, which shows that, as in the bosonic case,
field  redefinitions can be found that cast the action in a standard form
which is the superfield generalization of (\ref{generic}). A further 
generalization
also in the superfield context that relaxes the usual torsion
constraints has been given in \cite{Er:Ka:Ku:97}. However, this paper is mainly
concerned with an alternative formulation based on free differential algebras,
as will be explained in what follows. 

 The pure dilaton gravity 
part ({\it i.e.} in the absence of matter) of the CGHS and Jackiw-Teitelboim 
models can be given a gauge theoretic interpretation, 
in which the starting points are the two-dimensional extended Poincar\'e
and  de Sitter algebras respectively.  
  Given a Lie algebra $\cal G$ with commutators
\be
[T_a, T_b]= C_{ab}^c T_c \quad  ,           \label{algebra}
\ee
where $a=1,...,r=\hbox{dim}{\cal G}$, the gauge field
\be
A=A^a T_a
                                                     \label{gaugefield}
\ee
and a scalar $L$ with values in the dual of the above algebra, {\it i.e.}
\be
L=\eta_a T^a\quad , \quad T^a(T_b)=\delta^a_b               \label{scalar}
\ee
are introduced. Then a gauge invariant action for these fields is given by
\be
I=\int_{\cal M} L( F(A))= \int_{\cal M} \eta_a F^a\quad ,
                                                      \label{action}
\ee
where ${\cal M}$ is the two-dimensional spacetime
 and $F(A)$ is the  curvature of the connection $A$,
$F(A)=dA+A\wedge A$. 
Indeed,  under a gauge transformation of the connection $A$ the
curvature transforms under the adjoint representation of the superalgebra
(\ref{algebra}).  The action (\ref{action}) is then invariant under the gauge
transformations of the connection $A$ provided that $L$ transforms under 
 the coadjoint representation of the algebra. Note that the Lie algebra does 
not have to admit a non-degenerate invariant inner product but, if it does,
(\ref{action}) may be replaced by the integral of $<L,F>$, $<\ ,\ >$ being the
inner product symbol, where now $L$ is an {\it algebra}-valued quantity.
  The field equations of the theory are
\be
F(A)=0\quad ,\quad (dA^a=-\frac{1}{2}C^a_{bc}A^b
   \wedge A^c) \quad ,\quad d\eta_a+C_{ab}^cA^b\eta_c=0\quad .
                                                      \label{fieldeq}
\ee
   It is clear
from these that the scalar field $L$ is a Lagrange multiplier that imposes
the zero curvature condition for the connection $A$. Suitable choices for
Lie algebras lead to theories with action 
(\ref{action}) that can be interpreted as two-dimensional gravities. 

The CGHS model has been constructed out of two different Lie algebras. 
In \cite{Ver:92}, the Poincar\'e algebra with generators $M$ (Lorentz 
generator)
and  $\{P_a;\ a=1,2\}$ (translations), and commutators 
\be
[M, P_a]={\epsilon^b}_a P_b\quad ,\quad [P_a, P_b]=0\quad ,
                                                        \label{poincare}
\ee
was used. The Levi-Civit\`a symbol $\epsilon^{ab}$ is defined here by 
$\epsilon^{01}=1$, and the indices are raised and lowered using the 
Minkowski metric $\eta_{ab}$; the spacetime 
signature has been chosen to be $(-,+)$.
The same definition applies to the corresponding
Levi-Civit\`a symbol with space-time indices  $\epsilon^{\mu\nu}$,
 where $\epsilon_{01}$ is therefore equal to $-1$.
 The formulation based on (\ref{poincare}) had some problems that were naturally 
solved in \cite{Can:Jac:92} by starting from a central extension of this 
algebra instead.
This implies introducing a central generator $I$.  The non-vanishing
commutators of the new algebra are
\be
[M, P_a]={\epsilon^b}_a P_b,\qquad [P_a, P_b]=-\epsilon_{ab} I\ .
                                                        \label{central}
\ee
It was shown in \cite{Fuk:Kam:85} that the Jackiw-Teitelboim model may be 
formulated as a theory based on the de Sitter algebra with generators 
$\{M, P_a; a=1,2\}$ and commutators
\be
[M, P_a]={\epsilon^b}_a P_b,\qquad [P_a, P_b]=- \Lambda
\epsilon_{ab} M\ , 
                                                         \label{adS}
\ee
where $\Lambda$ is a constant. Supersymmetric extensions of these algebras have
been studied \cite{Can:Leb:94}, \cite{Le:Ri:95} and they lead, by using 
the $Z_2$-graded
version of the procedure just explained, to supergravity theories.  
In particular it is known that the algebra (\ref{poincare}) admits a (p,q) 
supersymmetric extension, the
algebra (\ref{central}) admits two different N=1 supersymmetric extensions and 
the algebra (\ref{adS}) admits a unique (1,1) extension.  

The above construction can be reinterpreted in the more general setting of 
free differential algebras (FDA). A free differential algebra \cite{Sul:77} 
generated
by the differential forms $G_i^{n_i}$, where $n_i$ is the degree of the form, 
is a mapping $G_i^{n_i}\mapsto dG_i^{n_i}$ defined by
\be
    dG_i^{n_i}=\sum_r\sum_{n_{j_1}+\dots +n_{j_r}=n_i+1\atop n_{j_1}\geq 1\dots
      n_{j_r}\geq 1}\alpha_i^{j_1\dots j_r}G_{j_1}^{n_{j_1}}\wedge\dots
    \wedge G_{j_r}^{n_{j_r}}  \quad ,                   \label{free}
\ee
where $\alpha_i^{j_1\dots j_r}$ are in general functions of the zero-forms 
($G_i^{n_i}$ such that $n_i=0$), in such a way that by virtue of Leibniz's rule
and the expressions for $dG_i^{n_i}$ themselves, $d(dG_i^{n_i})$ is identically
zero ({\it i.e.} vanishes without using any algebraic relation between the
forms $G_i^{n_i}$). Analogously, it is possible to define $Z_2$-graded
free differential algebras for which eq. (\ref{free}) remains the same,
the only difference being that now $G_i^{n_i}\wedge G_j^{n_j}=
(-1)^{n_in_j}(-1)^{\alpha_i\alpha_j}G_j^{n_j}\wedge G_i^{n_i}$, where the
extra factor $(-1)^{\alpha_i\alpha_j}$ (not present in the bosonic case)
takes into account the $Z_2$ parities of $G_i^{n_i}$ and $G_j^{n_j}$ ($\alpha_i$
and $\alpha_j$ respectively).
 As shown before, associated to every Lie algebra it is possible to
construct a FDA by demanding that the curvatures in (\ref{fieldeq}) vanish,
and the same can be said of superalgebras and $Z_2$-graded FDAs. 
Adding the E-L equations for the Lagrange multipliers, a larger FDA is obtained.
 From this
point of view, the equations of CGHS and Jackiw-Teitelboim models can be written 
as a FDA that has a subalgebra that corresponds to a finite-dimensional Lie 
algebra. For instance, eqs. (\ref{fieldeq}) in the Jackiw-Teitelboim case are
(the exterior product symbol $\wedge$ will be omitted from now 
on)
\be
  \eqalign{
   &de^a+{\epsilon^a}_b \omega e^b=0 \cr
    &d\omega-\frac{\Lambda}{2}\epsilon_{ab}e^ae^b=0\cr
    &d\eta+\eta_a {\epsilon^a}_b e^b=0\cr 
   &d\eta_a-\eta_b{\epsilon^b}_a\omega+\eta\Lambda \epsilon_{ab}e^b=0\cr
  }                                                    \label{JT}
\ee
where $A=e^ap_a+\omega M$, and $e^a=e^a_\mu dx^\mu$, $\omega=\omega_\mu dx^\mu$
give the zweibein and spin connection respectively.

One of the points of the article is to show that the family of models 
(\ref{generic}) (and its generalization to the locally supersymmetric case)
can be given a free differential algebraic formulation that does not
correspond in general to a finite-dimensional Lie (super)algebra. However,
the models still have an interpretation in terms of 
 symmetries and  as the dual 
of an {\it infinite} dimensional Lie (super)algebra. Moreover, this approach 
also provides an alternative method to obtain the locally supersymmetric 
generalization of (\ref{generic}). It is convenient to note here that
an approach to generic 2D dilatonic gravities different from
that of FDAs is provided by the Poisson sigma models of \cite{Kl:St:96}.

The other main point to be presented here is the following. Once the 
field equations of both dilatonic gravity and dilatonic supergravity are
written as free differential algebras containing one-forms and zero-forms,
it is possible to find their general solution in the differential form
language (for the solution of (\ref{generic}) in a specific gauge
see, for instance, \cite{Lou:Kun:94}). The search for the general solution 
without fixing the gauge may be motivated by the fact that in the bosonic case 
its knowledge  has been used to prove that the corresponding models
in the presence of conformal matter can be related by a 
canonical transformation to a  theory of free fields with the constraints 
of a certain string theory \cite{Cr:Iz:Na:Na:98}. It might happen
that something similar is possible in the context of supergravity 
theories and superstrings.

The organization of the paper is as follows. Section 2 is devoted to the 
bosonic case. There, the FDA corresponding to a generic dilatonic
gravity model will be given, and its group theoretical meaning will be 
explained. The section ends with the derivation of the general 
solution using the FDA structure of the equations. In section 3, the
locally supersymmetric case is studied. This will include the 
generalization of the bosonic FDA of the previous section, its relation
with the ordinary formulation, the general solution (which in this case 
has some peculiarities) and the coupling of the models to conformal
matter by using Noether's method. Finally, there is a section that contains
the conclusions and outlook.  
 
\section{The generic bosonic case}

The FDA given by 
\be
  \eqalign{ &de^a+{\epsilon^a}_b\omega e^b=0\cr
  & d\omega-\frac{V{'}}{2}e^ae^b\epsilon_{ab}=0\cr
 &d\eta+\eta_a{\epsilon^a}_b e^b=0\cr
  &d\eta_b+V\epsilon_{ab}e^a-{\epsilon^a}_b\omega\eta_a=0\cr}
                                                      \label{freeV}
\ee
obviously does not have a subalgebra that can be derived from a 
finite-dimensional Lie algebra when $V{'}(\eta)\neq$const., although 
generalizes (\ref{JT}) ($V=\Lambda\eta$ there). Furthermore, (\ref{freeV}) 
are the Euler-Lagrange (E-L) equations of the Lagrangian two-form 
\be
      L=\eta_a(de^a+{\epsilon^a}_b\omega e^b)+\eta d\omega
    -\frac{V}{2}\epsilon_{ab}e^ae^b\quad ,           \label{formV}
\ee
which is equivalent to the Lagrangian density (\ref{generic})
in the absence of matter. The equivalence
is proved by solving for $\omega$ and $\eta_a$,
\be
 \eqalign{
\eta_a&=-{\epsilon^b}_a e^\mu_b\partial_\mu\eta\quad ,\cr
\omega_\mu &=e^{-1}\epsilon^{\rho\nu}\partial\rho e^a_\nu e_{a\mu}
 \cr 
}                                                      \label{equivalence}
\ee
where $e=\det(e^a_\mu)=\sqrt{-g}$ ($g_{\mu\nu}=e_\mu^a e_\nu^b \eta_{ab}$),
and then substituting them into (\ref{formV}) to obtain an action that
only depends on $\eta$ and $e^a_\mu$. In doing so, the relations
$L=L_{\mu\nu}
dx^\mu \wedge dx^\nu=\epsilon^{\mu\nu}L_{\mu\nu}d^2x\equiv{\cal L}d^2x$
and $\epsilon^{\mu\nu}\partial_\mu
 \omega_\nu=eR$ are used. Actually, substituting some E-L equations
directly into the action does not necessarily mean that the 
equations obtained from the resulting action are the same as the ones
obtained by making the substitution in the remaining E-L equations.
However, this does happen in this case because  
the substituted fields ($\eta_a$, $\omega$) are precisely the
ones the equations of which are used ($\omega$ and $\eta_a$ 
respectively).
This fact also guarantees that $\omega$ and $\eta_a$ can be substituted into
the symmetry transformation laws.
 On the other hand, eqs. (\ref{freeV}) can be obtained from (\ref{JT})
 by letting $\Lambda$ depend on $\eta$ and demanding that the result
is still a FDA (this procedure will be explained in more detail in the
locally supersymmetric case).
 
The case $V=$constant deserves a comment. It corresponds to the CGHS
action, but it does not correspond to the {\it gauge formulation} of the
CGHS model given in \cite{Can:Jac:92} because, by equation (\ref{central}),
this formulation contains a field (the one corresponding to the central
generator $I$), which is not included in (\ref{freeV}).
 If $V$ is taken to be a constant in (\ref{freeV}), the free differential  
algebra obtained is one that does have a subalgebra dual to a Lie algebra
 (the one considered in \cite{Ver:92}), but that does not correspond
to a Lagrangian of the form (\ref{action}).

Although for a  general $V$ the FDA formulation is not the 
gauge theoretic formulation that corresponds
to a finite-dimensional Lie algebra, it is still possible to work
out the gauge symmetries of the FDA (\ref{freeV}). A way to do that is to
write $\delta e^a =dE^a+F^a$, $\delta \omega= d\Omega+G$, where $E^a$ and
$\Omega$ are the gauge parameters, and then fix both the a priori unknown
one-forms $F^a$, $G$, and  the zero forms
$\delta\eta$, $\delta\eta_a$ in such a way that 
the FDA is stable under the variations. The result is
\be
\eqalign{
\delta e^a &=dE^a-{\epsilon^a}_b\Omega e^b+
   {\epsilon^a}_b E^b\omega\cr
  \delta \omega&= d\Omega+V{'}\epsilon_{ab}E^ae^b\cr
  \delta\eta&=-\eta_a {\epsilon^a}_b E^b\cr
 \delta\eta_a&=-V\epsilon_{ba}E^b+{\epsilon^b}_a\Omega\eta_b
\quad .\cr
}                                                        \label{simfda}
\ee 
An interesting feature of these variations is that the action (\ref{formV})
is only quasi-invariant ({\it i.e.} invariant up to the differential of a 
one-form) when $V$ is not proportional to $\eta$.
In the case $V=\Lambda \eta$, these symmetries are the gauge transformations
of the connections associated to the de Sitter algebra, as can be seen
by computing the commutator of two such transformations. If this is done
in general, say, for two $E$ transformations, the following is 
obtained:
\be
  \eqalign{
   [\delta_{E},\delta_{E{'}}]e^a&= -{\epsilon^a}_be^b(V{'}
\epsilon_{cd}E^cE{'}^d)\quad ,\cr
  [\delta_{E},\delta_{E{'}}]\omega&=d(V{'}\epsilon_{ab}E^aE{'}^b)-
 V{''}\epsilon_{ab}E^aE{'}^b(d\eta+\eta_c{\epsilon^c}_de^d)
\quad . \cr
}                                                     \label{commua}
\ee 
Note that unless $V{''}=0$ the algebra only closes over the space
of solutions. Another characteristic that is due to the departure
from the gauge formulation is that, even when the field equations are
taken into account, 
\be
   [\delta_{E},\delta_{E{'}}]=\delta_{\Omega=V{'}\epsilon_{ab}E^aE{'}^b}
\quad ,                                               \label{commub}
\ee  
which means that, with the exception of the case $V=$const.,
the group of transformations that leave the FDA (\ref{freeV}) invariant is
intrinsically gauge (the commutator of two $E$ rigid transformations gives
a gauge $\Omega$ transformation). Alternatively, it may be said that
(\ref{commub}) reflects the fact that
 the ``structure constants" depend on the field $\eta$ and are therefore not 
constant. This can be seen by computing the (vector space) dual
of  (\ref{freeV}), for which the following elementary differential geometry
relations can be used: 
\be
    df(X)=X.f \quad ,\quad \alpha(X,Y)=X.\alpha(Y)-Y.\alpha(X)-\alpha([X,
 Y])\quad ,                                             \label{diffgeom}
\ee
where $\alpha$ is a one-form, $X$, $Y$ are vector fields and $f$ is a function.
Now, since $e^a(P_b)=\delta^a_b$, $e^a(M)=0=\omega(P_a)$ and $\omega(M)=1$,
(\ref{diffgeom}) lead to
\be
\eqalign{
  P_a(\eta)&=-\eta_b {\epsilon^b}_a\quad ,\quad 
    P_a(\eta_b)=-V\epsilon_{ab}\quad ,\cr
  M(\eta)&=0\quad ,\quad M(\eta_a)= {\epsilon^b}_a\eta_b\quad ,\cr
 [P_a,&P_b]=-V{'}\epsilon_{ab}M\quad ,\quad [M,P_a]=
  {\epsilon^b}_a P_b\quad .\cr
}                                                        \label{commuc}
\ee
The first two lines of (\ref{commuc}) mean that
$M$ and $P_a$ are the vector fields that generate the transformation
$\delta\eta$, $\delta\eta_a$ of (\ref{simfda}), which in the case $V{''}=0$ 
is the
coadjoint representation on the coalgebra of the corresponding Lie algebra.
The last line can be viewed as an infinite-dimensional Lie algebra one
of its generators is $V{'}M$. The commutator of this generator gives,
by virtue of Leibniz's  rule and (\ref{commuc}), new ones that are products 
of $M$ and $P_a$  by functions of $\eta$ and $\eta_a$. These in turn produce
new generators and so on. The end result is in general an algebra with an
infinite number of generators. The Lie algebras that arise here should not be
identified with the non-linear ones studied in \cite{Ike:94}.

An advantage of writing the field equations as in (\ref{freeV}) is that
the general solution can be easily obtained in the language of differential 
forms. The way to do it is to solve for  $\eta$ in terms of $\eta_a$
{\it instead} of solving for $\eta_a$ in terms
of $\eta$ (as it was done in (\ref{equivalence})), so that the solution depends 
on the  pair of free functions $\eta_a$. More explicitly, 
 from the equations for $d\eta$ and $d\eta_a$ it is easy to deduce that
\be
     \frac{1}{2}\eta^a\eta_a+J=C\quad ,                 \label{Eone}
\ee 
where $J(\eta)$ is defined by $J{'}=V$ and the constant $C$ is
related to the ADM energy \cite{Geg:Kun:Lou:95}. This expression gives 
implicitly $\eta$ in terms of $\eta_a$. Next, the equation for $d\eta_a$
can be rewritten as
\be
     e^a=\frac{1}{V}(\eta^a\omega-\epsilon^{ba}d\eta_b) \quad .
                                                        \label{eomega}
\ee
If this equation is then substituted into those of $de^a$ and $d\omega$,
{\it one} equation involving $\omega$ and $\eta_a$ is found:
\be
 d\left[\frac{\omega}{V}+\frac{1}{2}\left( \frac{1}{V[C-J]}+\frac{D}{C-J}\right)
     \epsilon^{ab}\eta_a d\eta_b\right]=0	\quad   ,       \label{domega}
\ee  
where $D$ is another constant. Equations (\ref{domega}), (\ref{eomega})
and (\ref{Eone}) constitute the solution of (\ref{freeV}),
of the form $\eta=\eta(\eta_a, g)$, $\omega=\omega(\eta_a,g)$ and 
$e^a=e^a(\eta_b,g)$, where $g$ is another free function which
comes from integration of (\ref{domega}). Therefore, the number of free
functions equals the number of gauge symmetries. 
 Of course, this solution is by construction consistent with 
(\ref{equivalence}). The possibility $V=0$
may be avoided by considering only functions of $\eta$ that
are always different from zero except possibly when
$\eta=0$ and excluding the points $x$ for which $\eta(x)=0$ from 
the spacetime manifold (see, for instance, \cite{Ku:Pe:Sh:98}). The presence of
$C-J$ in the denominator is related to the fact that the spacetime
points where it vanishes can usually be interpreted as a black hole
horizon.

\section{Generic 2D supergravities}

Before considering the locally supersymmetric case, it is necessary to 
fix some conventions, which are the following.
  Spinors are taken to be real and two dimensional
(they belong to the vector space of
 the reducible (1,1) spinorial representation of the Poincar\'e 
group in two dimensions); correspondingly the gamma matrices are real.
In terms of $\epsilon^{ab}$, the matrix $\gamma_3$ is given by $\gamma_3\equiv
\frac{1}{2}\epsilon_{ab}\gamma^a\gamma^b$. 
From this, it is possible to deduce some useful relations:
\be
   \epsilon^{\mu\nu}e_{b\mu}e_{c\nu}=e\epsilon_{bc}\quad ,\quad 
\epsilon_{ab}e^a_\mu e^b_\nu=e\epsilon_{\mu\nu}\quad ,\quad
\gamma_3\gamma_3=1 \quad ,\quad
\gamma^a \gamma_3={\epsilon^a}_b \gamma^b\quad ,              \label{use} 
\ee
and the Fierz reordering
\be
    \lambda{\bar\psi}=\frac{1}{2}{\bar\lambda}\gamma^b\psi\gamma_b-
     \frac{1}{2} {\bar\lambda}\psi
      +\frac{1}{2}{\bar\lambda}\gamma_3\psi\gamma_3           \label{fierz}
\ee
for any two spinors $\lambda$, $\psi$.
The following realization of the gamma matrices 
will be used (underlined indices are flat space indices):
\be
  \gamma^{\underline 0}=\left(\begin{array}{cr}0&-1\\ 1&0\end{array}
   \right)\quad ,\quad
  \gamma^{\underline 1}=\left(\begin{array}{cc}0&1\\ 1&0\end{array}\right)
\quad , \quad
  \gamma_3= \left(\begin{array}{cr}1&0\\ 0&-1\end{array}\right)\quad .
                                                       \label{matrices}
\ee

As announced in the introduction,
the generalization to the locally supersymmetric case of (\ref{freeV})
will be obtained from the FDA that leads to the supersymmetric
Jackiw-Teitelboim model, which is the graded de Sitter Lie superalgebra
OSP(1,1$\vert$1) \cite{Can:Leb:94}. The  field 
equations for the latter define the free differential algebra 
\be
    \eqalign{
      de^a&+ {\epsilon^a}_b\omega e^b+2i{\bar\psi}\gamma^a\psi=0\cr
     d\omega&-2m^2\epsilon_{ab}e^ae^b+4im{\bar\psi}\gamma_3\psi
    =0\cr
     d\psi&+\frac{1}{2}\omega\gamma_3\psi+m e^a\gamma_a\psi
     =0\cr
     d\eta_a&-\eta_b{\epsilon^b}_a\omega+4m^2e^b\epsilon_{ba}
     +im{\bar\chi}\gamma_a\psi=0\cr
      d\eta&+\eta_a{\epsilon^a}_b e^b+\frac{i}{2}{\bar\chi}\gamma_3
    \psi=0\cr
      d\chi &+m e^a\gamma_a\chi+\frac{1}{2}\omega\gamma_3\chi+
  4\eta_a\gamma^a\psi+8m\eta\gamma_3\psi=0 \quad ,\cr
      }                                               \label{osp} 
\ee
from which the Lie superalgebra can be immediately recovered by duality
($A=e^aP_a+\omega M+\psi^\alpha Q_\alpha$, and $Q_\alpha$ are the 
supersymmetry generators).
Next, the parameter $m$ is taken to be a function of $\eta$. If this is 
done in
the action the E-L equations of which are (\ref{osp}), a new term proportional 
to ${\bar\chi}e^a\gamma_a\psi$ has to appear in the equation for $d\omega$.
The result is, however, not a {\it free} differential algebra because
computing $dd$ for each form and equating the result to zero would give
extra algebraic relations between the forms. To convert the algebra into a
free one, the terms proportional to $m$ are substituted by terms of 
the same
form but multiplied by unknown functions of $\eta$. On the other hand, the
first equation of (\ref{osp}) should remain unaltered because it gives the usual
torsion corresponding to $\omega$. Then, imposing that $dde^a=0$ identically,
makes it necessary to add a term proportional to $\epsilon^{ab}e_ae_b\chi$
in the equation for $d\phi$, which has to come from another one in the 
Lagrangian two-form proportional to $\epsilon^{ab}e_ae_b{\bar\chi}\chi$.
This term introduces new ones in the equations of $d\eta_a$ and $d\omega$.
The requirement $dd=0$, when applied to the other equations, fixes the arbitrary 
functions of $\eta$ giving the result
\be
     \eqalign{
      de^a&+ {\epsilon^a}_b\omega e^b+2i{\bar\psi}\gamma^a\psi=0\cr
     d\omega&-2(uu{'}){'}\epsilon_{ab}e^ae^b+4iu{'}{\bar\psi}\gamma_3\psi
    +iu{''}{\bar\chi}e^a\gamma_a\psi+\frac{i}{16}u{'''}\epsilon^{ab}e_ae_b
     {\bar\chi}\chi=0\cr
     d\psi&+\frac{1}{2}\omega\gamma_3\psi+u{'}e^a\gamma_a\psi+\frac{1}{8}
      u{''}\epsilon^{ab}e_ae_b\chi=0\cr
     d\eta_a&-\eta_b{\epsilon^b}_a\omega+4uu{'}e^b\epsilon_{ba}
     +iu{'}{\bar\chi}\gamma_a\psi-\frac{i}{8}u{''}\epsilon_{ba}e^b{\bar\chi}
     \chi=0\cr
      d\eta&+\eta_a{\epsilon^a}_b e^b+\frac{i}{2}{\bar\chi}\gamma_3
    \psi=0\cr
      d\chi &+u{'}e^a\gamma_a\chi+\frac{1}{2}\omega\gamma_3\chi+4\eta_a\gamma^a
       \psi+8u\gamma_3\psi=0 \quad ,\cr
      }                                               \label{fda}
\ee
where $u(\eta)$ is an arbitrary function.
In fact, these relations are the Euler-Lagrange equations derived from the 
Lagrangian two-form
\be
  \eqalign{
   L&=\eta_a(de^a+{\epsilon^a}_b\omega e^b+2i{\bar\psi}\gamma^a\psi)
  +\eta d\omega-2uu{'}\epsilon_{ab}e^ae^b+4iu{\bar\psi}\gamma_3\psi\cr
    &+iu{'}{\bar\chi}e^a\gamma_a\psi+i{\bar\chi}(d\psi+\frac{1}{2}\omega
    \gamma_3\psi)+\frac{iu{''}}{16}\epsilon^{ab}e_ae_b{\bar\chi}\chi
    \quad .\cr
     }                                                \label{ltf}
\ee
Note that the Ads case is recovered when $u=m\eta$. It is now simple to 
look for the local supersymmetry transformations under which this algebra is 
invariant (the Lagrangian two-form is then 
in general quasi-invariant, as it happened in the bosonic case).
The comments made about the bosonic symmetries (\ref{simfda}), as well as the
way to obtain them, also apply
here.  The transformation rule for $\psi$ has to be of 
the form $\delta\psi=d\epsilon+\alpha$, where $\epsilon$ is the infinitesimal
parameter of the transformation and $\alpha$ is a one-form that can be
determined by substituting this expression into the third equation of 
(\ref{fda}),
and demanding that the terms containing $d\epsilon$ cancel. Moreover, this 
condition may be used to obtain the form of the variation for the remaining
one-forms. Having done that, it may be checked that the Lagrangian two-form
is indeed quasi-invariant under the variation, except when
$u=m\eta$, in which case it is strictly invariant. The following local
supersymmetry transformations are obtained:
\be
    \eqalign{
      \delta e^a&=4i{\bar\psi}\gamma^a\epsilon\cr
      \delta\psi&= d\epsilon+\frac{1}{2}\omega\gamma_3\epsilon+u{'}e^a\gamma_a
        \epsilon\cr
      \delta\omega &=-8iu{'}{\bar\epsilon}\gamma_3\psi-iu{''}{\bar\epsilon}
       \gamma^a\chi e_a\cr
      \delta \eta&= -\frac{i}{2}{\bar\chi}\gamma_3\epsilon\cr
       \delta\eta_a&=-iu{'}{\bar\chi}\gamma_a\epsilon\cr
        \delta\chi &=-8u\gamma_3\epsilon-4\eta_a\gamma^a\epsilon \quad .\cr
       }                                                    \label{susy}
\ee
An immediate consequence of (\ref{susy}) is that
the model with $u\propto\sqrt{\eta}$ is yet another supersymmetrization of
the CGHS model, different from the ones considered in \cite{Can:Leb:94}.
Specifically, the transformation rule for $\psi$ is $\delta\psi=
d\epsilon+\frac{1}{8}\omega\gamma_3\epsilon+c\frac{1}{\sqrt{\eta}}e^a
\gamma_a\epsilon$, where $c$ is a constant, in contrast with the two
cases of \cite{Can:Leb:94}, for which either there is no $e^a
\gamma_a\epsilon$ term or there is a term of the form
$(1-\gamma_3)e^a\gamma_a\epsilon$ times a constant.

As explained in the previous section, to
connect this formulation with the ordinary one the standard procedure is to 
write all the forms and scalars on space-time ($e^a=e^a_\mu dx^\mu$, 
$\omega=\omega_\mu dx^\mu$, $\psi=\psi_\mu dx^\mu$) and then to solve for 
$\eta_a$ in the fourth equation of (\ref{fda}) to obtain
\be
      \eta_a=-{\epsilon^b}_a e^\mu_b\partial_\mu\eta-\frac{i}{2}{\bar\chi}
   \gamma_3\psi_\mu e^\mu_b{\epsilon^b}_a\quad             \label{etaa}    
\ee
and use the first equation of (\ref{fda}) to determine
$\omega_\mu$ in terms of $e_\mu^a$ and $\psi_\mu$ as
\be
      \omega_\mu=e^{-1}\epsilon^{\rho\nu}\partial\rho e^a_\nu e_{a\mu}
   +2ie^{-1}\epsilon^{\rho\nu}{\bar\psi}_\rho\gamma_\mu\psi_\nu=
     {\bar \omega}_\mu+2ie^{-1}\epsilon^{\rho\nu}{\bar\psi}_\rho
    \gamma_\mu\psi_\nu                                     \label{connection}
\ee
where  ${\bar \omega}_\mu$ is the torsion-free spin connection. These two
equations are then used to  substitute $\eta_a$ and $\omega$ both into the 
action and into the local supersymmetry  transformations.  The ordinary
Lagrangian is obtained from the Lagrangian two-form as in the bosonic case,
and is given by
\be
     \eqalign{
    {\cal L}_{sg}&=eR\eta+4euu{'}+
     4iu\epsilon^{\mu\nu}{\bar\psi}_\mu\gamma_3\psi_\nu+iu{'}\epsilon^{\mu\nu}
    e^a_\mu{\bar\chi}\gamma_a\psi_\nu+i\epsilon^{\mu\nu}{\bar\chi}D_\mu\psi_\nu
    \cr
    &-\frac{ieu{''}}{8}{\bar\chi}\chi-4\eta ie(D_\mu{\bar\psi}^\mu\gamma_\nu
      \psi^\nu+{\bar\psi}^\mu\gamma_\nu D_\mu\psi^\nu)+2e{\bar\chi}\gamma_3
      \psi_\nu {\bar\psi}^\mu\gamma_\mu\psi^\nu\quad ,\cr
     }                                                        \label{lagrangian}
\ee 
where use is made of the fact that $\epsilon^{\mu\nu}\partial_\mu
{\bar \omega}_\nu=eR$, and
 $D_\mu$ is the covariant derivative with respect to the spin connection
${\bar \omega}_\mu$: $D_\mu\psi_\nu=\partial_\mu\psi_\nu+\frac{1}{2}
{\bar \omega}_\mu \gamma_3\psi_\nu$, and similarly for other spinors.
This is the locally supersymmetric version of (\ref{generic}), which
coincides, apart from conventions, with the superfield formulation of 
\cite{Par:Str:93}, once the appropriate field redefinitions are used to
get $K(\Phi)=0$ and $J(\Phi)=0$ there.
Note that, when $u{''}\neq 0$, it is possible to solve the algebraic equation 
for $\chi$, which gives a Lagrangian density independent of $\chi$.
The local supersymmetry transformations are
\be
      \eqalign{
        \delta e^a_\mu &=4i{\bar\psi}_\mu\gamma^a\epsilon\cr
     \delta\psi_\mu&=D_\mu\epsilon+u{'}e^a_\mu\gamma_a\epsilon-i{\bar\psi}_\rho
       \gamma_\mu\psi_\nu\gamma^{\rho\nu}\epsilon\cr
     \delta\eta&=-\frac{i}{2}{\bar\chi}\gamma_3\epsilon\cr
    \delta\chi&=-8u\gamma_3\epsilon+4\partial_\mu\eta\gamma^\mu\gamma_3\epsilon
  +2i{\bar\chi}\gamma_3\psi_\mu \gamma^\mu\gamma_3\epsilon\quad .\cr
   }                                                         \label{susya}
\ee

The general solution of the field equations written as a $Z_2$-graded
free differential algebra can be found by using the same procedure as in the
bosonic case, although there are a few differences. First, it is easy to
check, by using the last three equations of (\ref{fda}), that 
\be
\frac{1}{2}\eta^a\eta_a+2u^2-\frac{i}{8}u{'}{\bar\chi}\chi=C\quad ,
                                                                  \label{Etwo}
\ee
(cf. eq. (\ref{Eone})) where $C$ is a constant. This expression defines 
implicitly
$\eta$ in terms of $\eta^a$ and $\chi$. The last equation in (\ref{fda}) can be 
used to obtain $\psi$ in terms of $e^a$, $\omega$, $\eta^a$ and $\chi$. 
Multiplying it by $\eta_a\gamma^a+2u\gamma_3$ and then using (\ref{Etwo}),
the following equation is obtained:
\be
      (\eta_a\gamma^a+2u\gamma_3)\nabla\chi+(8C+iu{'}{\bar\chi}\chi)\psi=0
       \quad ,                                             \label{psia}
\ee 
where 
\be
    \nabla\chi:=d\chi+u{'}e^a\gamma_a\chi+\frac{1}{2}\omega\gamma_3\chi\quad .
                                                             \label{nabla}
\ee
In some cases, this equation does not determine $\psi$ because it is not
possible to divide by the factor $8C+iu{'}{\bar\chi}\chi$. This happens when
$C$ is a grassmann even number without an ordinary number part
({\it i.e.}, when the `body' of $C$ is zero), and when $C=0$.
The first case will not be considered because finding the solution implies
having to separate the components of $C$
in terms of its components in a basis of the Grassmann algebra, and the
use of non-commuting numbers is just a device motivated by the  
anticommuting character of the corresponding quantum operators and does not
have a physical meaning by itself. However, the $C=0$ case still has to be
considered, and it will be dealt with at the end of the section. In the 
other cases, eq. (\ref{psia}) can be solved without using the Grassmann algebra
structure, and the result can be substituted into the fourth equation of
(\ref{fda}), which in turn gives $e^a$ as an expression involving 
$\omega$, $\eta^a$ and $\chi$. Once the expressions for $\eta$, $\psi$ and 
$e^a$ are known, they can be substituted into the first three equations of
(\ref{fda}). They all give the same equation for $\omega$ (this is 
something that can be deduced from the integrability conditions of the 
last three equations of (\ref{fda})), so it is possible
to write the solution in terms of $\eta^a$ and $\chi$. The result is given
by (\ref{Etwo}) plus
\be
  \eqalign{
   \psi&=-\frac{1}{8C}\left[ 1-\frac{i}{8C}u{'}{\bar\chi}\chi\right]
  (\eta_a\gamma^a+2u\gamma_3)\nabla\chi\quad ,\cr
    e^a&=\left[4uu{'}-\frac{i}{8}u{''}{\bar\chi}\chi+\frac{i}{4C}(u{'})^2u
      {\bar\chi}\chi\right]^{-1}\left[\eta^a\omega
    \left( 1+\frac{i}{16C}u{'}{\bar\chi}\chi\right)-\epsilon^{ca}d\eta_c\right.
      \cr
    &\left.+\frac{iu{'}}{8C}\epsilon^{ca}{\bar\chi}\gamma_c(\eta_b\gamma^b+
      2u\gamma_3)d\chi\right]\quad ,\cr
    d&\left\{ \left[\frac{4}{uu{'}}
   +\frac{i}{8}\frac{u{''}}{(uu{'})^2}
     {\bar\chi}\chi\right]\omega\right.\cr
     &+\left[\left(\frac{2}{uu{'}[C-2u^2]}+
     \frac{D}{C-2u^2}\right)\left(1-\frac{iu{'}}{8[C-2u^2]}{\bar\chi}\chi
     \right)\right.\cr
    &\left.+\frac{i}{16u[C-2u^2]}\left(\frac{u{''}}{u(u{'})^2}-
     \frac{2}{C}+\frac{8u^2}{C^2}\right){\bar\chi}\chi\right]
     \epsilon^{ab}\eta_a d\eta_b\cr
     &+\left.i\left(\frac{1}{2Cu}-\frac{2u}{C^2}\right)
      {\bar\chi}\gamma_3 d\chi-\frac{i}{C^2}\eta_a{\bar\chi}\gamma^a d\chi
      \right\}=0\quad ,\cr
}                                                      \label{solutionb}
\ee
where $D$ is another constant. Note that there are five arbitrary functions:
$\eta^a$, $\chi$ and the one that comes from the integration of the
last equation in (\ref{solutionb}). This number coincides with the
number of gauge symmetries of the FDA.
 Of course, a gauge fixing greatly
simplifies this expression but, as stated before, it may be important
to control the gauge degrees of freedom.
\medskip

\noindent {\it The $C=0$ case}
\medskip

When $C=0$, eq. (\ref{psia}) may in principle be solved for the components
of $\psi$ in a basis of the underlying Grassmann algebra, although it is not 
completely determined due to the fact that the body of ${\bar\chi}\chi$
is zero. This is not desirable from the point of view of the 
quantum theory, so it is convenient to restrict the space of solutions to
those which can be expressed in terms of the fields themselves
and not their components. This means that to solve (\ref{psia})
${\bar\chi}\chi\psi$ has to vanish. 
Then it makes sense to try a solution of the form $\psi=\mu\chi$, where
$\mu$ is an one-form to be determined. Indeed, it can be checked, following
the same procedure as in the $C\neq 0$ case, that ($D$, $D{'}$ are constants)
\be
\eqalign{
\psi&=\frac{u{'}}{16u^2}e^a\eta_a\chi\quad ,\cr
 e^a&=\left(4uu{'}-\frac{i}{8}u{''}{\bar\chi}\chi\right)^{-1}(\eta^a
   \omega-\epsilon^{ba}d\eta_b)\quad ,\cr
  &\eta^a\eta_a+4u^2-\frac{i}{4}u{'}{\bar\chi}\chi=0\quad ,\cr
  &d\left\{\left[\frac{4}{uu{'}}+\frac{i}{8}\frac{u{''}}{(uu{'})^2}
{\bar\chi}\chi\right]\omega  -\left[
\frac{i}{u^2}\left(\frac{1}{32}\frac{u{''}}{(uu{'})^2}+D{'}
 \right){\bar\chi}\chi\right.\right.\cr
 &+\left.\left.\left( 1+\frac{i}{16}\frac{u{'}}{u^2}
 {\bar\chi}\chi\right)\left( \frac{1}{u^3u{'}}+\frac{D}{2u^2}\right)
 \right]\epsilon_{ab}\eta^ad\eta^b\right\}=0 \cr  
}                                                   \label{Czero}
\ee
together with $\nabla\chi=0$ and $\eta_a\gamma^a\chi+2u\gamma_3\chi=0$, 
provides a solution of the FDA (\ref{fda}) when $C=0$.

However, (\ref{Czero}) is not the only solution (this is not surprising, 
since it is expected that the solution includes two arbitrary functions 
to account for local supersymmetry). If $\psi,e^a,\omega,\eta$ is a solution,
the set $\psi{'},e^a,\omega,\eta$  with $\psi{'}=\psi+\sigma\chi+
{\bar\chi}\chi\zeta$ is also a solution provided
 $\sigma$ is a closed one-form and and $\zeta$ obeys 
$\eta_a\gamma^a\zeta+2u\gamma_3\zeta=0$ and $\nabla\zeta=0$.
 The final result is 
therefore (\ref{Czero}) except for the first equation, which takes the form
\be
   \psi=\frac{u{'}}{16u^2}e^a\eta_a\chi+\sigma\chi+{\bar\chi}\chi\zeta
        \quad .                                        \label{Czeroa}
\ee

Both $\chi$ and $\zeta$ have to satisfy the same system of equations,
with $e^a,\omega,\eta$ given in (\ref{Czero}). Explicitly, writing
$\xi$ to denote either the spinor or the spinorial one-form,
\be
   \eqalign{
  &\left(d+\frac{1}{2}\omega\gamma_3+u{'}e^a\gamma_a\right)\xi=0\quad ,\cr
   &(\eta_a\gamma^a+2u\gamma_3)\xi=0\quad .\cr   \label{Keq}
}
\ee
The solution given in (\ref{Czero}) is, as far as the bosonic fields is
concerned, equal to the solution of the theory without fermions plus some 
corrections proportional to ${\bar\chi}\chi$. These corrections are
irrelevant when solving (\ref{Keq}), due to the presence of ${\bar\chi}\chi$
multiplying the equations (\ref{Keq}) 
when $\xi=\zeta$ (see (\ref{Czeroa})) and to the presence of $\chi$ itself when
$\xi=\chi$, because expressions involving
the product of three $\chi$ spinors vanish.
 By virtue of the second and last equation in (\ref{susy}),
solving the equations is the same as, 
given a bosonic field configuration, looking for values of the
supersymmetry parameter $\epsilon(x)$ such that supersymmetry is preserved.
 These particular values of $\epsilon(x)$
receive the name of Killing spinors. Not every configuration admits 
Killing spinors, but it 
turns out that those with $C=0$ do, and in fact these are the only
configurations that  are solutions of the field
equations in the absence of matter and 
have Killing spinors (as can easily be seen by
computing the square of the matrix in the second equation of (\ref{Keq})).
To find the explicit expression of the solution of (\ref{Keq}),
it is convenient to define $\eta_{=\!\!\! \vert}=\eta_{\underline 0}
+\eta_{\underline 1}$, $\eta_{=}=\eta_{\underline 0}-\eta_{\underline 1}$
  (underlined indices correspond to flat space indices)
and write $\chi=\left(\begin{array}{c}\xi^R\\ \xi^L\end{array}\right)$
 in the basis corresponding to (\ref{matrices}). The solution may then be
written
\be
     d(\vert\eta_{=}\vert^{-1/2}\xi^R)=0\quad ,\quad 
    \xi^{L}=\frac{1}{2u}\eta_{=\!\!\!\vert}\; \xi^R\quad . \label{Killingb}
\ee
In this way, the closed form $\sigma$ of (\ref{Czeroa}) gives a free function, 
and the first equation of (\ref{Killingb}), when applied to $\zeta$, gives 
another one. Hence, the number of free functions is five as expected.

\medskip
{\it The coupling to conformal matter}
\medskip

When studying the physical consequences of 2D models, such as black hole
formation and evaporation, it is necessary to add matter fields. For this
 reason it may be interesting to do it in the supersymmetric case, providing 
explicit expressions in components. The next thing to do is therefore to couple 
these locally supersymmetric dilatonic gravity models to conformal 
matter. Here, this will be done by using the Noether method,
although the same result can be easily obtained using 
superfields. The starting point 
is the flat space, rigid (1,1) supersymmetry invariant matter Lagrangian 
\be
     {\cal L}_m=-\frac{1}{2} \eta^{\mu\nu}\partial_\mu f\partial_\nu f+
       \frac{i}{4}{\bar \lambda}\gamma^\mu \partial_\mu\lambda\quad , 
                                                             \label{rigid}
\ee 
where the first term corresponds in curved space to the usual conformal matter
Lagrangian, and the spinorial term makes it supersymmetric for the rigid
variation
\be
    \eqalign{\delta f&=i{\bar\epsilon}\lambda\cr
        \delta\lambda &=-2\partial_\mu f\gamma^\mu\epsilon\quad.\cr
    }                                                       \label{rigidv}
\ee
The curved space version of (\ref{rigid}) (which is the one needed to couple it 
to (\ref{lagrangian})) is obviously not invariant under the variations 
(\ref{rigidv}) when they are written in curved space and the
parameter $\epsilon$ is made space-time dependent. There are terms in the 
variation that come from the variation of $e^\mu_a$, and there are also terms 
proportional to $D_\mu\epsilon$ that would also appear even in flat space
because the variation is now local. The latter can be cancelled by adding to 
the action terms that involve $\psi_\mu$. They can be seen to be equal to
$\Delta_1{\cal L}= ie\partial_\mu f {\bar \psi}_\nu\gamma^\mu\gamma^\nu
\lambda$. Among the other terms, plus the new ones coming from the variation of 
$\Delta_1{\cal L}$, there are some that contain $D_\mu\lambda$. These can be 
cancelled by adding a new piece to the variation of $\lambda$: $\delta{'}\lambda
=2i{\bar \lambda}\psi_\mu\gamma^\mu\epsilon$. The new variation
contains terms 
involving $D_\mu\epsilon$, which means that the term $\Delta_2{\cal L}=
-\frac{e}{4}{\bar\psi}_\nu\gamma^\mu\gamma^\nu\psi_\mu
{\bar\lambda}\lambda$ must be added to the action. The process stops here, 
because at this point the complete variation vanishes up to a total derivative. 
The resulting Lagrangian density is then
\be
   {\cal L}={\cal L}_{sg}-\frac{1}{2} e g^{\mu\nu}\partial_\mu f\partial_\nu f+
       i\frac{e}{4}{\bar \lambda}\gamma^\mu D_\mu\lambda
      + ie\partial_\mu f {\bar \psi}_\nu\gamma^\mu\gamma^\nu
     \lambda -\frac{e}{4}{\bar\psi}_\nu\gamma^\mu\gamma^\nu\psi_\mu
      {\bar\lambda}\lambda \quad ,                            \label{total}
\ee
and the local variation of the matter fields is given by
\be
    \eqalign{\delta f&=i{\bar\epsilon}\lambda\cr
        \delta\lambda &=-2\partial_\mu f\gamma^\mu\epsilon
      +2i{\bar \lambda}\psi_\mu\gamma^\mu\epsilon \quad.\cr
    }                                                       \label{vmatter}
\ee

Due to the fact that the matter multiplet is the one corresponding to a 
conformally coupled matter field $f$, the coupling of matter to the locally
supersymmetric models is exactly the same as that for the pure 
Poincar\'e case. If a coupling of the form $eh(\eta)g^{\mu\nu}\partial_\mu f
\partial_\nu f$ was added (an interesting case, corresponding to a scalar
field in four dimensions is $V\propto \frac{1}{\sqrt{\eta}}$,
$h(\eta)\propto\eta$), both the variation of the matter fields and the 
matter Lagrangian density itself would have to involve the field $\chi$, 
because $\delta\eta =-\frac{i}{2}{\bar\chi}\gamma_3\epsilon$. However, it is
not immediate how to apply the Noether method in this case, and superspace
methods would presumably be more appropriate here.

\section{Conclusions and outlook}

This paper shows that a generic two-dimensional dilatonic gravity theory in 
the absence of matter can be expressed as a free differential algebra.
This has several consequences. First, there is room for a symmetry 
interpretation which generalizes that of the gauge theoretic formulation.
Second, it provides a method for obtaining the general solution in
terms of differential forms in both the bosonic and the locally supersymmetric
case. Third, it provides an alternative method to obtain the
generic supergravity Lagrangians.

It is still to be investigated how the free functions appearing in
the solutions obtained relate to the different gauge fixings. This will be 
important
when the programme of relating the dilatonic theories to free field
theories is carried out for the locally supersymmetric case. In that 
context, having a general solution of the starting models
in the absence of matter might help
to find the new canonical variables, or to prove that they exist.
Another point to be analyzed is whether it is possible to couple the
models to matter while maintaining the symmetries of the free differential 
algebras, as was done in \cite{Can:Jac:94} for the CGHS model.  

\medskip

The author would like to thank J.A. de Azc\' arraga, J. Negro,
J. Navarro-Salas, G. Papadopoulos and P.K. Townsend for useful discussions.

\end{document}